\documentclass[10pt]{iopart}
\usepackage{iopams}
\usepackage{hyperref}
\hypersetup{
colorlinks=true,
linkcolor=blue,
citecolor=blue,
urlcolor=blue}
\usepackage{color,soul}

\expandafter\let\csname equation*\endcsname\relax
\expandafter\let\csname endequation*\endcsname\relax

\usepackage{dsfont}

\usepackage{mathtools}
\DeclarePairedDelimiter\bra{\langle}{\rvert}
\DeclarePairedDelimiter\ket{\lvert}{\rangle}
\DeclarePairedDelimiterX\braket[2]{\langle}{\rangle}{#1 \delimsize\vert #2}
\begin{document}

\title[Phase space formulation of the quantum geometric tensor]
{Phase space formulation of the Abelian and non-Abelian quantum geometric tensor}

\author{Diego Gonzalez$^{1,2}$, Daniel Guti\'errez-Ruiz$^{1}$ and J. David Vergara$^{1}$}

\address{$^1$ Departamento de F\'isica de Altas Energ\'ias, Instituto de Ciencias Nucleares, Universidad Nacional Aut\'onoma de M\'exico, Apartado Postal 70-543, Ciudad de M\'exico, 04510, M\'exico}

\address{$^2$ Departamento de F\'isica, Cinvestav, Avenida Instituto Polit\'ecnico Nacional 2508, San Pedro Zacatenco, 07360, Gustavo A. Madero, Ciudad de M\'exico, M\'exico.}

\eads{\mailto{dgonzalez@fis.cinvestav.mx}, \mailto{daniel.gutierrez@correo.nucleares.unam.mx} and \mailto{vergara@nucleares.unam.mx}}

%- - - - - - - - - - - - - - - - - - - - - 

\begin{abstract}

The geometry of the parameter space is encoded by the quantum geometric tensor, which captures fundamental information about quantum states and contains both the quantum metric tensor and the curvature of the Berry connection. We present a formulation of the Berry connection and the quantum geometric tensor in the framework of the phase space or Wigner function formalism. This formulation is obtained through the direct application of the Weyl correspondence to the geometric structure under consideration. In particular, we show that the quantum metric tensor can be computed using only the Wigner functions, which opens an alternative way to experimentally measure the components of this tensor. We also address the non-Abelian generalization and obtain the phase space formulation of the Wilczek-Zee connection and the non-Abelian quantum geometric tensor. In this case, the non-Abelian quantum metric tensor involves only the non-diagonal Wigner functions. Then, we verify our approach with examples and apply it to a system of $N$ coupled harmonic oscillators, showing that the associated Berry connection vanishes and obtaining the analytic expression for the quantum metric tensor.  Our results indicate that the developed approach is well adapted to study the parameter space associated with quantum many-body systems.

\end{abstract}


\section{Introduction}

The last decade has seen remarkable advances in understanding the geometric aspects of the parameter space in several systems of condensed matter physics~\cite{SHI-JIAN2010,CAROLLO20201}. At the core of these developments is the quantum geometric tensor, which is a powerful tool to characterize the geometry of the eigenstates of Hamiltonians depending smoothly on external parameters. The symmetric part of this tensor gives rise to the quantum metric tensor on the system's parameter manifold~\cite{Provost1980}, whereas the antisymmetric part provides the curvature of the Berry connection, whose flux gives the (Berry) geometric phase~\cite{Berry45}. In turn, these geometrical objects have been used independently to reveal the appearance of quantum phase transitions~\cite{Chen2006,VenutiZanardi2007,Zanardi2007}.

One of the obstacles to study the quantum geometric tensor in its original (Hamiltonian) formulation is the difficulty in determining analytically or perturbatively the wave function. Encouraged by this fact and previous works in the context of the holographic principle~\cite{Miyaji2015,BAK2016200,Trivella_2017}, in Ref~\cite{Alvarez-Jimenez2017} was proposed a path integral formulation of the quantum geometric tensor, which turns out to be advantageous in quantum field theories or systems where the exact solution is not attainable~\cite{AlvarezVergara-2019}. In the framework of the phase space or Wigner function formalism there is also a formulation of the Berry curvature~\cite{Chruscinski2006}. This formulation sheds some light on the semiclassical relation between the Berry phase and the Hannay angle (see also Ref.~\cite{PATI1998178} for an early work); however, it was motivated mainly by a conjecture made by the author that has not been proved~\cite{Chruscinski2006}. 

The Wigner function formalism is not only a remarkable picture of quantum mechanics~\cite{Kim1991,zachos2005}, but also an appropriate framework for analyzing the correspondence between classical and quantum mechanics~\cite{Kenfack2004,Bernardini2019}. This picture is based on the Wigner function~\cite{WignerE1932}, which is a quasidistribution in phase space that has found applications in several branches of physics~\cite{Weinbub2018}. Furthermore, a substantial advantage of the Wigner function is that it can be measured experimentally~\cite{Smithey1993,Dunn1995}. This feature would be particularly important to allow, for instance, measurements of the quantum metric tensor. However, despite these attractive properties, there is no description of the quantum geometric tensor (or even of the quantum metric) in this formalism.

The purpose of this paper is to report a formulation of the Berry connection and the quantum geometric tensor in the framework of the Wigner function formalism. This formulation is achieved  using the Weyl correspondence~\cite{Weyl}  and involves, in addition to the Wigner functions, certain complex functions in phase space variables, which allow us to express these geometrical structures of the parameter space as integrals over phase space. Moreover, in this work we also provide the phase space formulation of the generalized non-Abelian quantum geometric tensor~\cite{YuQuan2010} and the non-Abelian Berry connection or Wilczek-Zee connection~\cite{Wilczek1984}, which are the corresponding geometrical structures that emerge in the case of a degenerate system. As in the Abelian case, the symmetric part of the non-Abelian quantum geometric tensor corresponds to a non-Abelian quantum metric tensor, while its antisymmetric part leads to a non-Abelian Berry curvature or Wilczek-Zee curvature. In particular,  we shall see that the Abelian and non-Abelian quantum metric tensors can be  computed using the Wigner functions purely, which opens a new line of attack on the problem of experimentally measuring the components of this tensor. To illustrate the use and validity of the new formulation, we present a well-known example of the literature, obtaining the expected results for the geometrical structures. As an application  of this approach, we study the parameter space of a system of $N$ coupled harmonic oscillators, showing that the associated Berry connection vanishes and obtaining the analytic expression for the quantum metric tensor.  Then, we consider the classical Hamiltonian corresponding to the above quantum system and obtain the classical analog~\cite{GGVmetric2019} of the resulting quantum metric tensor. We find that,  in the case where the transformation that diagonalizes the Hamiltonians does not depend on the system's parameters, the resulting quantum metric tensor and its classical analog are the same,  modulo the use of the Bohr-Sommerfeld quantization rule for action variables. We also apply this approach to a degenerate system composed of three coupled oscillators to determine the non-Abelian quantum metric tensor.

The outline of the paper is as follows. In Sec.~\ref{Framework}, we  present an overview of the fundamental structures underlying the parameter space and the Wigner-function formalism. In Sec.~\ref{formulation}, we derive the phase space formulation of the Abelian and non-Abelian quantum geometric tensor, as well as the formulation of the Berry connection and the Wilczek-Zee connection. In Sec.~\ref{examples}, we compute the Berry connection and the quantum geometric tensor for the generalized harmonic oscillator, showing how the phase space formulation leads to familiar results. In Sec.~\ref{Noscillators}, we apply the phase space approach to a system of $N$ coupled harmonic oscillators, which has been used in the study of entanglement  entropy~\cite{Bombelli1986,Srednicki1993}. In Sec.~\ref{NonAbelian}, we show how the phase space formulation works in the case of a degenerate system of three coupled oscillators.  Finally, in Sec.~\ref{sec:Conclusions} we present our conclusions.

\section{Framework}\label{Framework}

In order to set up the notations and the main assumptions, we begin this paper by briefly reviewing some aspects of the local geometry of the parameter space and the Wigner function formalism.  

\subsection{Geometry of the parameter space} 

Consider a quantum system defined by a Hamiltonian $\hat{\bf{H}}(\hat{\bf{q}},\hat{\bf{p}}; x)$, where $\hat{\bf{q}}=\{\hat{\bf{q}}_a\}$ and $\hat{\bf{p}}=\{\hat{\bf{p}}_a\}$ ($a,b,\dots\! =1,\dots,N$) are respectively the position and momentum operators, and $x(\tau)=\{x^i\}$ ($i,j,\dots\!=1,\dots,m$) is a set of $m$ real adiabatic parameters, i.e., slowly varying functions of time~$\tau$. To simplify notation, $x(\tau)$ will be denoted as $x$. If $\hat{\bf{H}}$ has at least one orthonormal eigenvector $\ket{n(x)}$ with nondegenerate eigenvalue $E_n(x)$, then the Abelian quantum geometric tensor defined in the parameter space of the system is given by
\begin{eqnarray}\label{QGT}
	Q^{(n)}_{ij} := \bra{\partial_i n}\left(1-\ket{n}\bra{n}\right) \ket{\partial_j n}, 
\end{eqnarray}
where $\partial_i = \partial/ \partial x^i$.  By construction, Eq~(\ref{QGT}) is invariant under the phase transformation~\cite{Provost1980}
\begin{eqnarray}\label{GaugeT}
	\ket{n(x)} \ \rightarrow \ \ket{n'(x)} = {\rm e}^{\rmi \alpha_n(x)} \ket{n(x)},
\end{eqnarray}
where $\alpha_n$ is an arbitrary real function of the parameters, meaning that we can choose the basis $\ket{n}$ up to a $U(1)$ gauge transformation (see also Ref.~\cite{Alvarez-Jimenez2016}). The relevance of this tensor lies in the fact that it provides the fundamental structures underlying the parameter space.

The (symmetric) real part of Eq.~(\ref{QGT}) gives the quantum metric tensor~\cite{Provost1980}
\begin{eqnarray}\label{QMT}
	g^{(n)}_{ij} = {\rm Re} \, Q^{(n)}_{ij},
\end{eqnarray}
which defines the line element as $dl^2 = g^{(n)}_{ij} \delta x^i \delta x^j$ and provides  a distance between the two neighbor states $\ket{n(x)}$ and $\ket{n(x+\delta x)}$ over the parameter space. Actually, the metric~(\ref{QMT}) stems from the leading term of the fidelity $f=| \braket{n(x)}{n(x+\delta x)}|$, a measurement of the ``similarity'' between the states $\ket{n(x)}$ and $\ket{n(x+\delta x)}$. 

The (antisymmetric) imaginary part of Eq.~(\ref{QGT}) yields the Berry curvature~\cite{Berry1985}
\begin{eqnarray}\label{BerryC}
	F^{(n)}_{ij}=-2 \, {\rm Im} \, Q^{(n)}_{ij}=\partial_i A^{(n)}_j-\partial_j A^{(n)}_i,
\end{eqnarray}
which defines the 2-form $F^{(n)}=\frac{1}{2}F^{(n)}_{ij} \delta x^i \wedge \delta x^j$ in the parameter space and is associated with  the (Abelian) Berry connection 
\begin{eqnarray}\label{BerryA}
	A^{(n)}_i := \rmi \braket{n}{\partial_i n}.
\end{eqnarray}
It is worthy of noticing that $A^{(n)}_i(x)$ is real since $\braket{ n}{\partial_i n}$ is purely imaginary and that, under the gauge transformation (\ref{GaugeT}), it changes as
\begin{eqnarray}\label{TransBerry}
	A^{(n)}_i \ \rightarrow \ A'^{(n)}_i=A^{(n)}_i-\partial_i\alpha_n,
\end{eqnarray}
which is the transformation law for a genuine $U(1)$ gauge connection. The integral of the 1-form $A^{(n)}=A^{(n)}_i\delta x^i $ over a closed path $C$ in the parameter space or the flux of the 2-form $F^{(n)}$ through a surface with boundary $C$ yields the gauge invariant Berry phase~\cite{Berry1985}.

On the other hand, if the Hamiltonian $\hat{\bf{H}}$ has a set of $g_n$ orthonormal eigenvectors $\ket{n_I(x)}$ ($I,J,\dots\!=1,2,\dots,g_n$) associated with the eigenvalue $E_n(x)$, i.e., $\hat{\bf{H}}(x)\ket{n_I(x)}=E_n(x)\ket{n_I(x)}$, then the non-Abelian quantum geometric tensor defined in the parameter space of the system is given by~\cite{YuQuan2010}
\begin{eqnarray}\label{nA-QGT}
	Q^{(n)}_{ijIJ} :=  \bra{\partial_i n_I} \left(1-\sum_{K=1}^{g_n} \ket{n_K}\bra{n_K}\right) \ket{\partial_j n_J}.
\end{eqnarray}
It is easy to see that in the nondegenerate case $g_n=1$, the non-Abelian tensor~(\ref{nA-QGT}) reduces to the Abelian one~(\ref{QGT}).  Furthermore, under the unitary transformation 
\begin{eqnarray}\label{nA-GaugeT}
	\ket{n_I(x)} \ \rightarrow \ \ket{n'_I(x)} = \sum_{J=1}^{g_n} \ket{n_J(x)} U_{JI}(x),
\end{eqnarray}
where $U_{IJ}(x)$ are the entries of a parameter-dependent unitary $g_n\times g_n$ matrix $U(x)$, the tensor (\ref{nA-QGT}) transforms covariantly 
\begin{eqnarray}\label{trans-nA-QGT}
	Q^{(n)}_{ijIJ} \ \rightarrow \ Q'^{(n)}_{ijIJ} = \sum_{K,L=1}^{g_n} U^*_{KI} Q^{(n)}_{ijKL} U_{LJ},
\end{eqnarray}
where `${}^*$' stands for complex conjugation.

By analogy with Eqs.~(\ref{QMT}) and~(\ref{BerryC}), the corresponding (symmetric) non-Abelian quantum metric tensor and the (antisymmetric) non-Abelian Berry curvature or Wilczek-Zee curvature are given by
\begin{eqnarray}\label{nA-QMT}
	g^{(n)}_{ijIJ} = \frac12(Q^{(n)}_{ijIJ}+Q^{(n)*}_{ijJI}),
\end{eqnarray}
and 
\begin{eqnarray}\label{nA-BerryF}
	F^{(n)}_{ijIJ}=\rmi (Q^{(n)}_{ijIJ}-Q^{(n)*}_{ijJI}),
\end{eqnarray}
respectively. Moreover, the associated non-Abelian Berry connection or Wilczek-Zee connection is defined by~\cite{Wilczek1984,Wilczek1989book}
\begin{eqnarray}\label{nA-BerryC}
	A^{(n)}_{iIJ} := \rmi \braket{n_I}{\partial_i n_J}. 
\end{eqnarray}
This connection provides the entries of a $g_n\times g_n$  Hermitian matrix for each $i$ and, under the transformation (\ref{nA-GaugeT}), transforms as a proper gauge potential
\begin{eqnarray}\label{trans-nA-A}
	A^{(n)}_{iIJ} \ \rightarrow \ A'^{(n)}_{iIJ} = \sum_{K,L=1}^{g_n} U^*_{KI} A^{(n)}_{iKL} U_{LJ} + \rmi \sum_{K=1}^{g_n} U^*_{KI} \partial_i U_{KJ}.
\end{eqnarray}

\subsection{Wigner-function formalism} 

Given an operator $\hat{\bf{Q}}(\hat{\bf{q}},\hat{\bf{p}}; x)$, the Weyl correspondence  associates a quantum phase space function $\mathcal{Q}(q,p ; x)$ defined as~\cite{Weyl} 
\begin{eqnarray}\label{WeylT}
	\mathcal{Q}(q,p;x)=\int_{-\infty}^{\infty} {\rm d}^Ny \, {\rm e}^{ -\frac{{\rm i}p\cdot y}{\hbar}} \bra{q+\tfrac{y}{2}} \hat{\bf{Q}}(\hat{\bf{q}},\hat{\bf{p}}; x)\ket{q-\tfrac{y}{2}},
\end{eqnarray}
which is known as the Weyl transform of $\hat{\bf{Q}}$. Here,  the variables $q=\{q_a\}$ and $p=\{p_a\}$ are respectively the eigenvalues of the operators $\hat{\bf{q}}$ and $\hat{\bf{p}}$, and we have written ${\rm d}^N \!y\equiv{\rm d}y_1\dotsi{\rm d}y_N$, and $p\cdot y\equiv\sum_{a=1}^{N} p_a y_a$. An important property of the Weyl correspondence is that it allows to write the trace of the product of two operators, $\hat{\bf{Q}}$ and $\hat{\bf{O}}$, as
\begin{eqnarray}\label{WeylTrace}
	{\rm Tr} (\hat{\bf{Q}}\hat{\bf{O}})=\frac{1}{(2 \pi \hbar)^N} \int_{-\infty}^{\infty}  {\rm d}^Nq \, {\rm d}^Np\,\mathcal{Q} \mathcal{O},
\end{eqnarray}
where $\mathcal{Q}$ and  $\mathcal{O}$ are the Weyl transform of these operators.

The Wigner function $W_n(q,p;x)$, which is the main tool of this formalism, is defined as the function corresponding to the density operator $\hat{\rho}(x)$. More precisely, for a pure state, $\hat{\rho}_n(x)=\ket{n(x)}\bra{n(x)}$, it is given by~\cite{WignerE1932}
\begin{eqnarray}\label{WignerF}
	W_n(q,p;x)	=\frac{1}{(2\pi \hbar)^N} \int_{-\infty}^{\infty} {\rm d}^Ny \,  {\rm e}^{ -\frac{{\rm i}p\cdot y}{\hbar}} \psi_{n}(q+\tfrac{y}{2};x)\psi^*_{n}(q-\tfrac{y}{2};x),
\end{eqnarray}
where $\psi_{n}(q+\tfrac{y}{2};x)=\braket{q+\tfrac{y}{2}}{n(x)}$ and $\psi^*_{n}(q-\tfrac{y}{2};x)=\braket{n(x)}{q-\tfrac{y}{2}}$. Thus, the Wigner function provides a phase space representation of the quantum state $\ket{n(x)}$. 

From Eq.~(\ref{WeylTrace}) with $\hat{\bf{Q}}\rightarrow\hat{\rho}_n$, it is straightforward to see that the expectation value of an operator $ \hat{\bf{O}}$ can be written as
\begin{eqnarray}\label{WignerAvg}
	\langle \hat{\bf{O}} \rangle_n={\rm Tr} (\hat{\rho}_n\, \hat{\bf{O}})=\int_{-\infty}^{\infty}  {\rm d}^Nq \, {\rm d}^Np\, W_n \, \mathcal{O}.
\end{eqnarray}
This expression is reminiscent of the average of the function $\mathcal{O}$ (to which $\hat{\bf{O}}$ corresponds) with respect to the phase space ``quasiprobability'' distribution $W_n$, which is real but not always positive. Notice that setting  $\hat{\bf{O}}$ equal to the identity operator $\hat{\mathds{1}}$ in Eq.~(\ref{WignerAvg}) and using the fact that the Weyl transform of $\hat{\mathds{1}}$ is $1$, we have
\begin{eqnarray}\label{norma}
	\int_{-\infty}^{\infty}  {\rm d}^Nq \, {\rm d}^Np\, W_n=1,
\end{eqnarray}
which shows that $W_n$ is normalized in phase space. For later purposes, we also write down the following expression derived by taking $\hat{\bf{O}}\rightarrow\hat{\rho}_n$ in Eq.~(\ref{WignerAvg}): 
\begin{eqnarray}\label{norma2}
	\int_{-\infty}^{\infty}  {\rm d}^Nq \, {\rm d}^Np\, W_n^2=\frac{1}{(2\pi \hbar)^N}.
\end{eqnarray}
For more details on the Wigner function, we refer the reader to Ref.~\cite{Case2008}.

\section{Phase space formulation of the geometry of the parameter space}\label{formulation}

This section is devoted to recasting the geometrical structures involved in the parameter space into the Wigner function formalism; both the Abelian and non-Abelian cases are addressed. The approach we will follow here takes advantage of the Weyl correspondence and is  different from that used in Ref.~\cite{Chruscinski2006}, where the Abelian and non-Abelian Berry curvatures are considered. In essence, the idea is very straightforward. It relies on the observation that these geometrical structures can be expressed as the trace of quantum operators, which will allow us to apply Eq.~(\ref{WeylTrace}), and then link with the phase space formalism.

\subsection{Abelian case}

Let us first consider the Abelian Berry connection. We start by writing Eq.~(\ref{BerryA}) as 
\begin{eqnarray}\label{TrA}
	A_i^{(n)}={\rm Tr}\, \hat{\bf{A}}^{(n)}_i,
\end{eqnarray}
where $\hat{\bf{A}}^{(n)}_i$ is a quantum operator defined by
\begin{eqnarray}\label{operatorA}
	\hat{\bf{A}}^{(n)}_i:= \rmi \ket{\partial_i n}\bra{n}.  
\end{eqnarray}
Note that $\hat{\bf{A}}^{(n)}_i$ is non-Hermitian, and then its Weyl transform
\begin{eqnarray}\label{WeylT-A}
	\mathcal{A}_i^{(n)}(q,p;x)&=&\rmi\int_{-\infty}^{\infty} {\rm d}^Ny \, {\rm e}^{ -\frac{{\rm i}p\cdot y}{\hbar}} \braket{q+\tfrac{y}{2}}{\partial_i n} \braket{n}{q-\tfrac{y}{2}} \nonumber\\
	&=&\rmi\int_{-\infty}^{\infty} {\rm d}^Ny \, {\rm e}^{ -\frac{{\rm i}p\cdot y}{\hbar}} \, \partial_i \psi_{n}(q\!+\!\tfrac{y}{2};x) \,\psi^*_{n}(q\!-\!\tfrac{y}{2};x),
\end{eqnarray}
is a complex function in phase space variables. Actually, by using Eqs.~(\ref{WignerF}) and~(\ref{WeylT-A}), it is not hard to show that
\begin{eqnarray}\label{ImA}
	{\rm Im} (\mathcal{A}_i^{(n)})=\frac{(2\pi \hbar)^N}{2} \partial_i W_n.
\end{eqnarray}

Having Eq.~(\ref{TrA}), we can now apply Eq.~(\ref{WeylTrace}) to relate the phase space function $\mathcal{A}_i^{(n)}$ with the Berry connection. Indeed, taking  $\hat{\bf{Q}}\rightarrow\hat{\mathds{1}}$ and $\hat{\bf{O}}\rightarrow\hat{\bf{A}}^{(n)}_i$ in Eq.~(\ref{WeylTrace}) and using Eq.~(\ref{TrA}), we have
\begin{eqnarray}\label{Wigner-BerryConn}
	A^{(n)}_i=\frac{1}{(2 \pi \hbar)^N}\int_{-\infty}^{\infty}  {\rm d}^Nq \, {\rm d}^Np\, \mathcal{A}_i^{(n)}.
\end{eqnarray}
This equation provides an expression of the Berry connection in the Wigner-function formalism. It can be checked that, under the gauge transformation~(\ref{GaugeT}), the function $\mathcal{A}_i^{(n)}$ transforms according to
\begin{eqnarray}\label{TransAtilde}
	\mathcal{A}_i^{(n)} \ \rightarrow \ \mathcal{A}{'}_i^{(n)}=\mathcal{A}_i^{(n)}-(2\pi \hbar)^N W_n \, \partial_i\alpha_n,
\end{eqnarray}
from which, together with Eq~(\ref{norma}), it follows that the connection (\ref{Wigner-BerryConn}) satisfies the transformation law (\ref{TransBerry}), as expected. Notice that the transformation law for $\mathcal{A}_i^{(n)}$ is similar to the one of a $U(1)$  gauge connection, but not identical because of the presence of $W_n$ in Eq.~(\ref{TransAtilde}). This is a consequence of the fact that, while the gauge transformation~(\ref{GaugeT}) is restricted to the parameter space, $\mathcal{A}_i^{(n)}(q,p;x)$ is a mixed structure of the parameter space and the physical phase space. Of course,  it is possible to consider a  less restrictive and more general gauge transformation (see Ref.~\cite{Alvarez-Jimenez2016}, for instance). It is also worth noting that only the real part of the phase space function $\mathcal{A}_i^{(n)}$ contributes to the Berry connection~(\ref{Wigner-BerryConn}), which is a consequence  of Eq.~(\ref{ImA}) and the fact that
\begin{eqnarray}\label{dWigner}
	\int_{-\infty}^{\infty}  {\rm d}^Nq \, {\rm d}^Np\, \partial_i W_n=0.
\end{eqnarray}

To recast the Abelian quantum metric tensor into the Wigner-function formalism,  here we follow the same procedure as in the case of the Berry connection. We observe that the quantum geometric tensor~(\ref{QGT}) can be written as 
\begin{eqnarray}\label{TrQ}
	Q^{(n)}_{ij} ={\rm Tr}\, \hat{\bf{Q}}^{(n)}_{ij},
\end{eqnarray}
where $\hat{\bf{Q}}^{(n)}_{ij}$ is an operator defined by
\begin{eqnarray}\label{operatorQ}
	\hat{\bf{Q}}^{(n)}_{ij}:=(\hat{\bf{A}}^{(n)\dagger}_i-\hat{\bf{A}}^{(n)}_i)\hat{\bf{A}}^{(n)}_j,
\end{eqnarray}
with $\hat{\bf{A}}^{(n)}_i$ given by Eq.~ (\ref{operatorA}). Therefore, noting that $\hat{\bf{A}}^{(n)\dagger}_i-\hat{\bf{A}}^{(n)}_i=-\rmi \partial_i \hat{\rho}_n$, from Eq.~(\ref{WeylTrace}) with $\hat{\bf{Q}}\rightarrow-\rmi \partial_i \hat{\rho}_n$ and $\hat{\bf{O}}\rightarrow\hat{\bf{A}}^{(n)}_j$ it follows that
\begin{eqnarray}\label{Wigner-QGTW}
	Q^{(n)}_{ij}=-\rmi \int_{-\infty}^{\infty}  {\rm d}^Nq \, {\rm d}^Np\, \partial_i W_n \, \mathcal{A}_j^{(n)}, 
\end{eqnarray}
which, after using Eq.~(\ref{ImA}), becomes
\begin{eqnarray}\label{Wigner-QGT}
	Q^{(n)}_{ij}=-\frac{2\rmi}{(2\pi \hbar)^N} \int_{-\infty}^{\infty}  {\rm d}^Nq \, {\rm d}^Np\, {\rm Im} (\mathcal{A}_i^{(n)}) \, \mathcal{A}_j^{(n)}.
\end{eqnarray}
This is a formulation of the quantum metric tensor within the phase space formalism. Notice that we only need the knowledge of  $\mathcal{A}_i^{(n)}$ in order to compute the quantum geometric tensor~(\ref{Wigner-QGT}). This shows that the new phase space functions $\mathcal{A}_i^{(n)}$,  introduced here for calculating the Berry connection, actually  encode all the relevant information embodied in the parameter space. On the other hand, it can be checked, by using Eq.~(\ref{TransAtilde}) together with  
\begin{eqnarray}\label{dWigner2}
	\int_{-\infty}^{\infty}  {\rm d}^Nq \, {\rm d}^Np\, W_n \partial_i W_n=0,
\end{eqnarray}
which follows from Eq.~(\ref{norma2}), that Eq.~(\ref{Wigner-QGT}) is gauge invariant under the transformation~(\ref{GaugeT}).

Separating Eq.~(\ref{Wigner-QGT}) into its real and imaginary parts, we obtain the expression for Abelian quantum metric tensor
\begin{eqnarray}\label{qmt-wigner}
	g^{(n)}_{ij}=\frac{2}{(2\pi \hbar)^N} \int_{-\infty}^{\infty}  {\rm d}^Nq \, {\rm d}^Np\,  {\rm Im} (\mathcal{A}_i^{(n)}) \, {\rm Im} (\mathcal{A}_j^{(n)}),
\end{eqnarray}
which, after using Eq.~(\ref{ImA}), can be written as 
\begin{eqnarray}\label{qmt-wignerW}
	g^{(n)}_{ij}=\frac{(2\pi \hbar)^N}{2} \int_{-\infty}^{\infty}  {\rm d}^Nq \, {\rm d}^Np\,  \partial_i W_n \, \partial_j W_n,
\end{eqnarray}
and the expression for the Abelian Berry curvature tensor
\begin{eqnarray}\label{BerryC-wigner}
	F^{(n)}_{ij}=\frac{4}{(2\pi \hbar)^N} \int_{-\infty}^{\infty}  {\rm d}^Nq \, {\rm d}^Np\,  {\rm Im} (\mathcal{A}_i^{(n)}) \, {\rm Re} (\mathcal{A}_j^{(n)}).
\end{eqnarray}

Some remarks are in order. First, note that in contrast to the standard approach to determine the Abelian quantum metric tensor (namely, Eq.~(\ref{QMT})), the phase space approach based on Eq.~(\ref{qmt-wignerW}) does not require the knowledge of wave functions. This is because the Wigner functions can also be obtained as the solutions of suitable functional equations in phase space, bypassing the use of wave functions~\cite{zachos2005}. Second, note that the expression~(\ref{BerryC-wigner}) for the Abelian Berry curvature is different from the one conjectured in Ref.~\cite{Chruscinski2006}, although both expressions must yield identical results. In fact, while Eq.~(\ref{BerryC-wigner}) only involves the phase space function $\mathcal{A}_{i}^{(n)}$, the expression proposed in~\cite{Chruscinski2006}, which uses Eq.~(\ref{WignerAvg}) and takes elements of the curvature associated with the classical Hannay angle~\cite{Hannay1985}, requires the knowledge of variables $q=\{q_a\}$ and $p=\{p_a\}$ in terms of the angle-action variables. In this regard, it is worth recalling that the angle-action variables are restricted to classical integrable systems, and hence the expression of Ref.~\cite{Chruscinski2006} can be applied only to those quantum systems whose classical counterpart is integrable, in contrast to Eq~(\ref{BerryC-wigner})  which does not have that limitation.  Furthermore, we would also like to point out that Eqs.~(\ref{Wigner-BerryConn}) and~(\ref{qmt-wigner})--(\ref{BerryC-wigner}) for the Abelian geometrical structures have not, to our knowledge, been reported in the literature. They are new and  can be useful not only to gain more insight in the nature of the parameter space of quantum systems, but also for practical applications in a variety of quantum models. Third, notice that in Eq.~(\ref{TrA})  we could also consider the operator $\hat{\bf{B}}^{(n)}_i:=(\beta_1 \hat{\bf{A}}^{(n)}_i+\beta_2 \hat{\bf{A}}^{(n)\dagger}_i)/(\beta_1+\beta_2)$ with $\beta_1$ and $\beta_2$ being arbitrary complex numbers, because of $A_i^{(n)}={\rm Tr}\, \hat{\bf{B}}^{(n)}_i$. Nevertheless, a more complicated choice could be inconvenient for practical applications and does not give any further insight into the nature of the parameter space.

\subsection{Non-Abelian case}

Here, we extend the phase space description of the parameter space to include the non-Abelian geometrical structures. Following the procedure considered for the Abelian case, we begin by expressing the Wilczek-Zee connection as
\begin{eqnarray}\label{nA-TrA-Wigner}
	A^{(n)}_{iIJ} = {\rm Tr}\, \hat{\bf{A}}^{(n)}_{iIJ},
\end{eqnarray}
where now the associated quantum operator takes the form
\begin{eqnarray}\label{nA-operatorA}
	\hat{\bf{A}}^{(n)}_{iIJ}:= \rmi \ket{\partial_i n_J}\bra{n_I}.  
\end{eqnarray}
The Weyl transform for this operator is given by
\begin{eqnarray}\label{WeylT-nA}
	\mathcal{A}_{iIJ}^{(n)}(q,p;x)&=&\rmi\int_{-\infty}^{\infty}  {\rm d}^Ny \, {\rm e}^{ -\frac{{\rm i}p\cdot y}{\hbar}} \braket{q+\tfrac{y}{2}}{\partial_i n_J} \braket{n_I}{q-\tfrac{y}{2}} \nonumber\\
	&=&\rmi\int_{-\infty}^{\infty}  {\rm d}^Ny \, {\rm e}^{ -\frac{{\rm i}p\cdot y}{\hbar}}  \partial_i \psi_{nJ}(q+\tfrac{y}{2};x) \psi^*_{nI}(q-\tfrac{y}{2};x),
\end{eqnarray}
and, as in the Abelian case, it gives rise to a complex function in phase space variables, since $\mathcal{A}^{(n)}_{iIJ}$  is a non-Hermitian operator. 

By combining Eq.~(\ref{nA-TrA-Wigner}) and Eq.~(\ref{WeylTrace}) with $\hat{\bf{Q}}\rightarrow\hat{\mathds{1}}$ and $\hat{\bf{O}}\rightarrow\hat{\bf{A}}^{(n)}_{iIJ}$, it is direct to see that the expression for the Wilczek-Zee connection in phase space formalism is
\begin{eqnarray}\label{Wigner-nABerryConn}
	A^{(n)}_{iIJ}=\frac{1}{(2 \pi \hbar)^N}\int_{-\infty}^{\infty}  {\rm d}^Nq \, {\rm d}^Np\, \mathcal{A}_{iIJ}^{(n)},
\end{eqnarray}
which is the natural generalization of the (Abelian) Berry connection~(\ref{Wigner-BerryConn}). It is worth noting that Eq.~(\ref{Wigner-nABerryConn}) gives the entries of a $g_n\times g_n$  Hermitian matrix for each $i$, as it should be. This can be proved as follows. From Eq.~(\ref{WeylT-nA}) we have
\begin{eqnarray}\label{nA-Atildecong}
	\mathcal{A}_{iIJ}^{(n)}=\mathcal{A}_{iJI}^{(n)*} + \rmi (2 \pi \hbar)^N \, \partial_i W_{nJI},
\end{eqnarray}
where $W_{nIJ}$ are the non-diagonal Wigner functions 
\begin{eqnarray}\label{nA-f}
	W_{nIJ}:=\frac{1}{(2\pi \hbar)^N} \int_{-\infty}^{\infty} {\rm d}^Ny \,  {\rm e}^{ -\frac{{\rm i}p\cdot y}{\hbar}} \psi_{nI}(q+\tfrac{y}{2};x)\psi^*_{nJ}(q-\tfrac{y}{2};x).
\end{eqnarray}
Then, plugging Eq.~(\ref{nA-Atildecong}) into Eq.~(\ref{Wigner-nABerryConn}) and using the fact that
\begin{eqnarray}\label{nA-Wigner-int}
	\int_{-\infty}^{\infty}  {\rm d}^Nq \, {\rm d}^Np\, W_{nIJ}=\delta_{IJ},
\end{eqnarray}
it is straightforward to see that $A^{(n)}_{iIJ}=A^{(n)*}_{iJI}$.

Let us now prove that the connection (\ref{Wigner-nABerryConn}) obeys the gauge transformation law~(\ref{trans-nA-A}) for the Wilczek-Zee connection. Under the unitary transformation~(\ref{nA-GaugeT}), the phase space function $\mathcal{A}_{iIJ}^{(n)}$ changes as
\begin{eqnarray}\label{trans-nA-Atilde}
 \fl	\mathcal{A}^{(n)}_{iIJ} \ \rightarrow \ \mathcal{A}'^{(n)}_{iIJ} =\sum_{K,L=1}^{g_n} \left[ U^*_{KI} \mathcal{A}^{(n)}_{iKL} U_{LJ}+ \rmi (2 \pi \hbar)^N  \, U^*_{KI} \, W_{nLK} \, \partial_i U_{LJ}\right],
\end{eqnarray}
which generalizes Eq.~(\ref{TransAtilde}). Note that while the second term on the right side of Eq.~(\ref{TransAtilde}) is purely imaginary, the corresponding term in Eq.~(\ref{trans-nA-Atilde}) is complex in general.  Now, taking into account Eq~(\ref{trans-nA-Atilde}), Eq~(\ref{Wigner-nABerryConn}) immediately implies
\begin{eqnarray}
\fl	A^{(n)}_{iIJ} \ \rightarrow \ A'^{(n)}_{iIJ} = \sum_{K,L=1}^{g_n} \left[U^*_{KI} A^{(n)}_{iKL} U_{LJ} + \rmi \, U^*_{KI} \partial_i U_{LJ}   \int_{-\infty}^{\infty}  {\rm d}^Nq \, {\rm d}^Np\, W_{nLK}  \, \right],
\end{eqnarray}
which, after using Eq.~(\ref{nA-Wigner-int}), becomes the  transformation law~(\ref{trans-nA-A}).

We now turn to the phase space  formulation of the non-Abelian quantum geometric tensor. We find that Eq.~(\ref{nA-QGT}) can be written as
\begin{eqnarray}\label{nA-QGT-Weyl}
	Q^{(n)}_{ijIJ} ={\rm Tr}\, \hat{\bf{Q}}^{(n)}_{ijIJ},
\end{eqnarray}
where the operator $\hat{\bf{Q}}^{(n)}_{ijIJ}$ is defined as
\begin{eqnarray}
	\hat{\bf{Q}}^{(n)}_{ijIJ}:=-\rmi \, \partial_i \hat{\mathfrak{P}}_n \, \hat{\bf{A}}^{(n)}_{jIJ}.
\end{eqnarray}
with $\hat{\mathfrak{P}}_n:=\sum_{I=1}^{g_n} \ket{n_I}\bra{n_I}$ the projection operator. Then, taking $\hat{\bf{Q}}\rightarrow-\rmi \, \partial_i \hat{\mathfrak{P}}_n$ and $\hat{\bf{O}}\rightarrow\hat{\bf{A}}^{(n)}_{jIJ}$ in Eq.~(\ref{WeylTrace}), it is clear that Eq~(\ref{nA-QGT-Weyl}) is equivalent to
\begin{eqnarray}\label{Wigner-nA-QGTW}
	Q^{(n)}_{ijIJ}=-\rmi \int_{-\infty}^{\infty}  {\rm d}^Nq \, {\rm d}^Np\, \sum_{K=1}^{g_n} \partial_i W_{nKK} \, \mathcal{A}_{jIJ}^{(n)}, 
\end{eqnarray}
Let us write this equation directly in terms of the phase space function $\mathcal{A}_{iIJ}^{(n)}$. Using Eqs.~(\ref{WeylT-nA}) and (\ref{nA-f}), it can be demonstrated that 
\begin{eqnarray}\label{nA-ImA}
	{\rm Im} (\mathcal{A}_{iKK}^{(n)})=\frac{(2\pi \hbar)^N}{2} \partial_i W_{nKK}, 
\end{eqnarray}
which is the analog of Eq.~(\ref{ImA}). Thus,
substituting Eq.~(\ref{nA-ImA}) into Eq.~(\ref{Wigner-nA-QGTW}), we get
\begin{eqnarray}\label{Wigner-nA-QGT}
	Q^{(n)}_{ijIJ}=\frac{-2\rmi}{(2\pi \hbar)^N} \int_{-\infty}^{\infty}  {\rm d}^Nq \, {\rm d}^Np\, \sum_{K=1}^{g_n} {\rm Im} (\mathcal{A}_{iKK}^{(n)}) \, \mathcal{A}_{jIJ}^{(n)},
\end{eqnarray}
which  is an expression for the non-Abelian quantum geometric tensor in the phase space formalism. This equation tells us that, as in the Abelian case, all we need to know in order to compute the non-Abelian quantum geometric tensor is the phase space function $\mathcal{A}_{iIJ}^{(n)}$.

Now we are in a position to find the expressions for the non-Abelian quantum metric tensor and Wilczek-Zee curvature. From Eqs.~(\ref{nA-QMT}) and (\ref{Wigner-nA-QGT}), the non-Abelian quantum metric tensor turns out to be
\begin{eqnarray}\label{Wigner-nA-QMT}
	g^{(n)}_{ijIJ}=\frac{-\rmi}{(2\pi \hbar)^N} \int_{-\infty}^{\infty} {\rm d}^Nq \, {\rm d}^Np \sum_{K=1}^{g_n}  {\rm Im} (\mathcal{A}_{iKK}^{(n)})  (\mathcal{A}_{jIJ}^{(n)}-\mathcal{A}_{jJI}^{(n)*}),
\end{eqnarray}
which, with the help of Eqs.~(\ref{nA-Atildecong}) and (\ref{nA-ImA}), takes the form
\begin{eqnarray}\label{Wigner-nA-QMT2}
	g^{(n)}_{ijIJ}=\frac{(2\pi \hbar)^N}{2}  \int_{-\infty}^{\infty}  {\rm d}^Nq \, {\rm d}^Np \sum_{K=1}^{g_n} \partial_i W_{nKK} \, \partial_j W_{nIJ},
\end{eqnarray}
whereas from Eqs.~(\ref{nA-BerryF}) and (\ref{Wigner-nA-QGT}), the Wilczek-Zee  curvature is recasted as
\begin{eqnarray}\label{Wigner-nA-BerryC}
	F^{(n)}_{ijIJ}=\frac{2}{(2\pi \hbar)^N} \int_{-\infty}^{\infty} {\rm d}^Nq \, {\rm d}^Np  \sum_{K=1}^{g_n}  {\rm Im} (\mathcal{A}_{iKK}^{(n)})  (\mathcal{A}_{jIJ}^{(n)}+\mathcal{A}_{jJI}^{(n)*}).
\end{eqnarray}
We remark here that, as in the Abelian case, the expression~(\ref{Wigner-nA-BerryC}) for the Wilczek-Zee curvature is different from the one proposed in  Ref.~\cite{Chruscinski2006}.  In fact, whereas Eq.~(\ref{Wigner-nA-BerryC}) only depends on the phase space function $\mathcal{A}_{iIJ}^{(n)}$, the expression of Ref.~\cite{Chruscinski2006} requires the non-diagonal Wigner functions and the variables $q=\{q_a\}$ and $p=\{p_a\}$ in terms of the angle-action variables, meaning that it is restricted to quantum systems that have a classical integrable counterpar. It is also worth mentioning that the phase space formulation obtained in this subsection for the non-Abelian geometrical structures is new and can be used to deal with a wide variety of degenerate quantum systems.

The following sections are devoted to exemplify and apply this phase space approach to the parameter space.

\section{Illustrative example}\label{examples}

Let us take the archetypal system of a generalized harmonic oscillator, which is defined by the quantum Hamiltonian
\begin{eqnarray}\label{gho:quantumH}
	\hat{\bf{H}}=\frac{1}{2}\left[X\hat{\bf{q}}^{2}+Y(\hat{\bf{q}}\hat{\bf{p}}+\hat{\bf{p}}\hat{\bf{q}})+Z\hat{\bf{p}}^{2}\right],
\end{eqnarray}
where $x=\{x^{i}\}=(X,Y,Z)$ ($i,j,\dots\!=1,2,3$) are the adiabatic parameters. The corresponding normalized wave functions  are well known and take the form 
\begin{eqnarray}\label{gho:wavefunctgen}
	\psi_{n}(q;x)=\left(\frac{\omega}{Z\hbar}\right)^{1/4}\chi_{n}\left(q\sqrt{\frac{\omega}{Z\hbar}}\right)\exp\left(-\frac{iYq^{2}}{2Z\hbar}\right),
\end{eqnarray}
where $n$ are non-negative integers, $\omega= \sqrt{XZ-Y^{2}}$ is the parameter-dependent frequency which requires $XZ-Y^{2}>0$, and $\chi_{n}(\xi)=\left(2^{n}n!\sqrt{\pi}\right)^{-1/2}{\rm e}^{-\xi^{2}/2}H_{n}(\xi)$ are the Hermite functions, with $H_{n}(\xi)=(-1)^n {\rm e}^{\xi^2} \frac{d^n}{d\xi^n}{\rm e}^{-\xi^2}$ being the Hermite polynomials. 

Our first task is to obtain the associated phase space function  $\mathcal{A}_i^{(n)}$. Plugging Eq.~(\ref{gho:wavefunctgen}) into Eq.~(\ref{WeylT-A}) and using the relations 
\begin{subequations}
	\begin{eqnarray}
		&&\frac{d \chi_{n}}{d\xi}  = \sqrt{\frac{n}{2}} \ \chi_{n-1} - \sqrt{\frac{n+1}{2}} \ \chi_{n+1}, \label{Hermite1}\\
		&&\xi \, \chi_{n} = \sqrt{\frac{n}{2}} \ \chi_{n-1} + \sqrt{\frac{n+1}{2}} \ \chi_{n+1}, \label{Hermite2}
	\end{eqnarray}
\end{subequations}
it is straightforward to see that $\mathcal{A}_i^{(n)}$ is given by
\begin{eqnarray}\label{gho:Atilde}
	\mathcal{A}_i^{(n)}=\frac{\pi \hbar Z}{2 \omega}  \left\{ \rmi  \partial_i \left( \frac{\omega}{Z} \right)  \Xi^{(-)}_{n} + \partial_i \left( \frac{Y}{Z} \right)  \left[\Xi^{(+)}_{n} + (2n+1) W_n \right] \right\},
\end{eqnarray}
where $W_n$ is the corresponding  Wigner function~\cite{Chruscinski2006},
\begin{eqnarray}\label{gho:Wigner}
	W_n=\frac{(-1)^n}{\pi \hbar} {\rm e}^{-\lambda/2} L_{n}(\lambda), \qquad (\lambda:=4H/\hbar\omega),
\end{eqnarray}
with $H=(1/2)\left(Xq^2+2Ypq + Zp^2\right)$ and $L_n(\lambda)$ the Laguerre polynomials. Furthermore, we have defined the phase space functions
\begin{eqnarray}\label{gho:Xi}
	\Xi^{(\pm)}_{n}:=\sqrt{n(n-1)} f_{n-2,n} \pm \sqrt{(n+1)(n+2)} f_{n+2,n},
\end{eqnarray}
where 
\begin{eqnarray}\label{nondiagonalWigner}
	f_{n\pm 2,n}=\frac{1}{2\pi \hbar} \int_{-\infty}^{\infty} {\rm d}y \,  {\rm e}^{ -\frac{{\rm i}p\cdot y}{\hbar}} \psi_{n\pm 2}(q+\tfrac{y}{2};x)\psi^*_{n}(q-\tfrac{y}{2};x),
\end{eqnarray}
for $n-2\geq0$, are the non-diagonal Wigner functions, which explicitly read
\begin{subequations}
	\begin{eqnarray}
		f_{n-2,n}&=& \frac{(-1)^{n+1} 2\left(P-\rmi \omega Q\right)^2}{\sqrt{n(n-1)}\, \pi \hbar^2  \omega} {\rm e}^{-\frac{\lambda}{2}} L^{(2)}_{n-2}(\lambda), \\
		f_{n+2,n}&=& \frac{(-1)^{n+1} 2 \left(P+\rmi \omega Q\right)^2}{\sqrt{(n+1)(n+2)}\, \pi \hbar^2  \omega} {\rm e}^{-\frac{\lambda}{2}} L^{(2)}_{n}(\lambda).
	\end{eqnarray}
\end{subequations}
Here, $L^{(\alpha)}_n(\lambda)$ are the associated Laguerre polynomials ($L^{(0)}_n(\lambda)=L_n(\lambda)$) and we have used the transformation
\begin{eqnarray}
	Q=\frac{1}{\sqrt{Z}}q, \qquad P=\sqrt{Z}\left( p +\frac{Y}{Z} q \right).
\end{eqnarray}

With this at hand,  Eq.~(\ref{Wigner-BerryConn}) can now be readily applied to Eq.~(\ref{gho:Atilde}). By doing so, and using Eq.~(\ref{norma}) together with the fact that~\cite{zachos2005}
\begin{eqnarray}\label{Intf}
	\int_{-\infty}^{\infty}  {\rm d}q \, {\rm d}p\, f_{n\pm 2,n}=0,
\end{eqnarray}
we arrive at the Berry connection 
\begin{eqnarray}\label{gho:Bconnetion}
\fl	A_{1}^{(n)}(x)=0,\qquad A_{2}^{(n)}(x)=\left(n+ \frac12\right)\frac{1}{2\omega}, \qquad A_{3}^{(n)}(x)=-\left(n+ \frac12\right)\frac{Y}{2Z\omega}. 
\end{eqnarray}
This result coincides with the usual expression found in the literature (see Ref~\cite{GGVmetric2019}, for instance). Notice that as a consequence of Eq.~(\ref{Intf}), only the term involving $W_n$ in Eq.~(\ref{gho:Atilde}) contributes to the non-zero Berry connection~(\ref{gho:Bconnetion}). This means that for the standard Harmonic oscillator ($Y=0$), in whose case Eq.~(\ref{gho:Atilde}) reduces to $ \mathcal{A}_i^{(n)}=\frac{\rmi \pi \hbar Z}{2 \omega} \Xi^{(-)}_{n} \partial_i \sqrt{\frac{X}{Z}}$, the resulting Berry connection is zero, as expected.  Regarding the Berry curvature, it can be computed directly from Eq.~(\ref{BerryC}) with Eq.~(\ref{gho:Bconnetion}) and the result is the well-known expression
\begin{eqnarray}\label{gho:Bcurvature}
	&&F_{12}^{(n)}(x)=-\left(n+ \frac12\right)\frac{Z}{4\omega^{3}},\qquad 
	F_{13}^{(n)}(x)=\left(n+ \frac12\right)\frac{Y}{4\omega^{3}}, \nonumber\\	&&F_{23}^{(n)}(x)=-\left(n+ \frac12\right)\frac{X}{4\omega^{3}}.
\end{eqnarray}

We now focus on the computation of the quantum metric tensor, which can be done either from Eq.~(\ref{qmt-wigner}) with the phase space function~(\ref{gho:Atilde}) or from Eq.~(\ref{qmt-wignerW}) with the Wigner function~(\ref{gho:Wigner}). Indeed, taking the imaginary part of Eq.~(\ref{gho:Atilde}) and using the relation 
\begin{eqnarray}
	L_n^{(2)}-L_{n-2}^{(2)}=L_n-2\frac{\partial L_n}{\partial \lambda},
\end{eqnarray}
or directly calculating the partial derivatives of Eq~(\ref{gho:Wigner}) with respect to the parameters, it follows that
\begin{eqnarray}\label{gho:dW}
	(\partial_i W_n)_{q,p} &=& \frac{1}{\pi \hbar}{\rm Im}(\mathcal{A}_i^{(n)}) \nonumber \\
	&=&-\frac{(-1)^n}{2\pi \hbar} {\rm e}^{-\lambda/2}\,\left(L_n-2\frac{\partial L_n}{\partial \lambda} \right) (\partial_i\lambda)_{q,p}.
\end{eqnarray}
Then, substituting Eq.~(\ref{gho:dW}) into Eq.~(\ref{qmt-wignerW}), and taking into account
\begin{eqnarray}\label{intLagaguerre}
	\int_{0}^{\infty}  {\rm d}\lambda\, {\rm e}^{-\lambda} \, \lambda^2\,\left(L_n(\lambda)-2\frac{\partial L_n(\lambda)}{\partial \lambda} \right)^2  = 2 (n^2+n+1),
\end{eqnarray}
we arrive at
\begin{eqnarray}\label{gho:QIM}
	g^{(n)}_{ij}(x)=\frac{n^{2}+n+1}{32\omega^{4}}\left(\begin{array}{ccc}
		Z^{2} & -2YZ & 2Y^{2}-XZ\\
		-2YZ & 4XZ & -2XY\\
		2Y^{2}-XZ & -2XY & X^{2}
	\end{array}\right),
\end{eqnarray}	
which is the same expression for the quantum metric tensor as that obtained in Ref.~\cite{GGVmetric2019} by using Eq.~(\ref{QMT}) with the wave function (\ref{gho:wavefunctgen}). This corroborates that the phase space formulation of the quantum metric tensor, namely Eq.~(\ref{qmt-wignerW}), yields the right results. 

\section{$N$ coupled harmonic oscillators}\label{Noscillators}

We now want to go further and extend our analysis of the parameter space to the case of $N$ coupled harmonic oscillators. Specifically, the system under consideration is described by the Hamiltonian
\begin{eqnarray}\label{Nosc:quaH}
	\hat{\bf{H}}=\frac{1}{2} \sum_{a=1}^{N} \hat{\bf{p}}_a^2 + \frac{1}{2} \sum_{a,b=1}^{N} K_{ab} \hat{\bf{q}}_a  \hat{\bf{q}}_b ,
\end{eqnarray}
where $K_{ab}(=K_{ba})$ are the entries of an $N\times N$ symmetric and positive-definite matrix $K(x)$, which is constructed out of the adiabatic parameters $x=\{x^{i}\}$ ($i,j,\dots\!=1,\dots,n$). This system has been employed to study the properties of entanglement  entropy~\cite{Bombelli1986,Srednicki1993}, with remarkable consequences.

It is convenient to start by introducing the linear transformation 
\begin{eqnarray}\label{Nosc:transf}
	\hat{\bf{Q}}_a=\sum_{b=1}^{N}  U_{ab} \hat{\bf{q}}_b, \qquad  \hat{\bf{P}}_a=\sum_{b=1}^{N}  U_{ab} \hat{\bf{p}}_b,
\end{eqnarray}
where $U_{ab}$ are the entries of an $N\times N$ orthogonal matrix $U$ such that $K=U^T\Omega^2 U$, with $\Omega={\rm diag}(\omega_1,\dots,\omega_N)$ being a diagonal matrix whose elements $\omega_a$ are the frequencies of the system. Using Eq.~(\ref{Nosc:transf}), the Hamiltonian (\ref{Nosc:quaH}) becomes
\begin{eqnarray}\label{Nosc:quaH2}
	\hat{\bf{H}}=\frac{1}{2} \sum_{a=1}^{N} \left( \hat{\bf{P}}_a^2 + \omega_a^2 \hat{\bf{Q}}_a^2 \right),
\end{eqnarray}
which enables us to write the associated normalized wave function as 
\begin{eqnarray}\label{Nosc:wave}
	\psi_{n_1,\dots,n_N}(q_1,\dots,q_N;x)=\prod_{a=1}^{N} \psi_{n_a}(Q_a;x),
\end{eqnarray}
where $\psi_{n_a}(Q_a;x)$ is the wave function of the $a$th uncoupled oscillator with quantum number $n_a$:
\begin{eqnarray}\label{sco:wavea}
\psi_{n_a}(Q_a;x)=\left(\frac{\omega_a}{\hbar}\right)^{1/4}\chi_{n_a}\left(Q_a\sqrt{\frac{\omega_a}{\hbar}}\right).
\end{eqnarray}
Here we restrict ourselves to the case where the eigenvalues 
\begin{eqnarray}\label{Nsco:Eigenval}
	E_{n_1,\dots,n_N}=\sum_{a=1}^{N}\left(n_a+\frac12\right)\hbar \omega_{a},
\end{eqnarray}	
are nondegenerate, which ensures that we stay in the Abelian setting.

With the wave function at hand, we can apply Eq.~(\ref{WeylT-A}) to get the corresponding phase space function $\mathcal{A}_i^{(n_1,\cdots,n_N)}$. Indeed, taking into account Eqs.~(\ref{Hermite1}) and (\ref{Hermite2}), and making the change of variables 
\begin{eqnarray}\label{Nsco:transfY}
	Y_{a}=\sum_{b=1}^{N}  U_{ab} \, y_{b},
\end{eqnarray}
it is relatively straightforward to show that Eq.~(\ref{WeylT-A}) with Eq~(\ref{Nosc:wave}) leads to
\begin{eqnarray}\label{Nsco:Atilde}
	\mathcal{A}_i^{(n_1,\dots,n_N)}&=&\rmi (2\pi \hbar)^N  \Bigg(\frac{1}{4} \sum_{a=1}^{N} \frac{\partial_i \omega_{a}}{\omega_{a}} \, \Xi_{n_a}  \prod_{\substack{b=1, \\ b\neq a}}^{N} W_{n_b}  \nonumber \\
	&&+\sum_{a,b,c=1}^{N} \sqrt{\frac{\omega_{a}}{\omega_{c}}} \, \partial_iU_{ab} \, U_{cb} \, \Theta^{(+)}_{n_a} \Theta^{(-)}_{n_c} \prod_{\substack{d=1, \\ d\neq a,c}}^{N} W_{n_d} \Bigg),
\end{eqnarray}
where $W_{n_a}$ are phase space functions given by
\begin{eqnarray}\label{sco:Wigner}
W_{n_a}=\frac{(-1)^{n_a}}{\pi \hbar} {\rm e}^{-\frac{\lambda_a}{2}} L_{n_a}(\lambda_a), \qquad (\lambda_a:=4H_a/\hbar\omega_a),
\end{eqnarray}
with $H_a=(1/2)(P_{a}^{2}  + \omega_{a}^{2} Q_{a}^{2} )$ and so that
\begin{eqnarray}\label{Nsco:wigner}
	W_{n_1,\dots,n_N}(q_1,\dots,q_N,p_1,\dots,q_N;x)=\prod_{a=1}^{N} W_{n_a}(Q_a,P_a;x),
\end{eqnarray}
is the Wigner function of the system. Furthermore, the functions $\Xi_{n_a}$ are defined by
\begin{eqnarray}\label{sco:Xi}
\Xi_{n_a}:=\sqrt{n_a(n_a-1)} f_{n_a-2,n_a}- \sqrt{(n_a+1)(n_a+2)} f_{n_a+2,n_a},  
\end{eqnarray}
with the non-diagonal Wigner functions
\begin{eqnarray}\label{nondiagonalWigner2}
\fl f_{n_a\pm 2,n_a}:=\frac{1}{2\pi \hbar} \int_{-\infty}^{\infty} {\rm d}Y_a \,  {\rm e}^{ -\frac{{\rm i}P_a Y_a}{\hbar}} \psi_{n_a\pm 2}(Q_a+\tfrac{Y_a}{2};x) \times \psi_{n_a}(Q_a-\tfrac{Y_a}{2};x),
\end{eqnarray}
which explicitly read 
\begin{subequations}
	\begin{eqnarray}
	&&f_{n_a-2,n_a}=  \frac{(-1)^{n_a+1} 2 \, (P_a-\rmi \omega_a Q_a)^2}{\sqrt{n_a(n_a-1)} \,\pi \hbar^2  \omega_a}  {\rm e}^{\frac{-\lambda_a}{2}} L^{(2)}_{n_a-2}(\lambda_a), \\
	&&f_{n_a+2,n_a}=  \frac{(-1)^{n_a+1} 2\, (P_a+\rmi \omega_a Q_a)^2}{\sqrt{(n_a+1)(n_a+2)}\, \pi \hbar^2  \omega_a} {\rm e}^{\frac{-\lambda_a}{2}}  L^{(2)}_{n_a}(\lambda_a),
	\end{eqnarray}
\end{subequations}
whereas the functions $\Theta^{(\pm)}_{n_a}$ have the form
\begin{eqnarray}\label{Nsco:Xi}
	\Theta^{(\pm)}_{n_a}=\sqrt{\frac{n_a}{2}} \, f_{n_a-1,n_a} \pm \sqrt{\frac{n_a+1}{2}} f_{n_a+1,n_a},  
\end{eqnarray}
where $f_{n_a\pm1,n_a}$ have the same definition as in Eq.~(\ref{nondiagonalWigner2}), but with $n_a\pm2$ replaced by $n_a\pm1$, and explicitly read 
\begin{subequations}
	\begin{eqnarray}
		&&f_{n_a-1,n_a}= \frac{(-1)^{n_a+1} 2^{1/2}  (\rmi P_a+ \omega_a Q_a)}{ n_a^{1/2} \pi \hbar^{3/2}  \omega_a^{1/2}}  {\rm e}^{\frac{-\lambda_a}{2}} L^{(1)}_{n_a-1}(\lambda_a), \\
		&&f_{n_a+1,n_a}= \frac{(-1)^{n_a+1} 2^{1/2} (\rmi P_a-\omega_a Q_a)}{(n_a+1)^{1/2} \, \pi \hbar^{3/2}  \omega_a^{1/2}} {\rm e}^{\frac{-\lambda_a}{2}} L^{(1)}_{n_a}(\lambda_a).
	\end{eqnarray}
\end{subequations}

It should be noted that for systems where the matrix $U$ does not depend on the parameters ($\partial_i U_{ab}=0$), the terms of the second line of Eq.~(\ref{Nsco:Atilde}) vanish and hence the function $\mathcal{A}_i^{(n_1,\dots,n_N)}$ reduces to $\mathcal{A}_i^{(n_1,\dots,n_N)}=  \frac{\rmi (2\pi \hbar)^N}{4} \sum_{a=1}^{N} \frac{\partial_i \omega_{a}}{\omega_{a}} \, \Xi_{n_a}  \prod_{\substack{b=1, \\ b\neq a}}^{N} W_{n_b}$. This is the case, for instance, of the system of two symmetric coupled oscillators, which is described by the Hamiltonian 
\begin{eqnarray}\label{sco:Hamil}
\hat{\bf{H}}=\frac{1}{2} \left[ \hat{\bf{p}}_{1}^{2}+\hat{\bf{p}}_{2}^{2}+k (\hat{\bf{q}}_{1}^{2}+\hat{\bf{q}}_{2}^{2})+k^{\prime} (\hat{\bf{q}}_{1}-\hat{\bf{q}}_{2})^{2}  \right],
\end{eqnarray}
with the adiabatic parameters $\{x^{i}\}=(k,k^{\prime})$, and is of interest in the study of  quantum entanglement~\cite{Srednicki1993,Chandran2019} and circuit complexity in quantum field theory~\cite{Jefferson2017}. Indeed, for such a system the entries of the matrix $U$ are given by  $U_{11}=U_{12}=U_{21}=-U_{22}=1/\sqrt{2}$ and the function $\mathcal{A}_i^{(n_1,n_2)}$ takes the simple form $\mathcal{A}_i^{(n_1,n_2)}=\rmi (\pi \hbar)^2  \left( \frac{\partial_i \omega_{1}}{\omega_{1}} \Xi_{n_1} W_{n_2}  + \frac{\partial_i \omega_{2}}{\omega_{2}} \Xi_{n_2} W_{n_1}  \right)$ where $\omega_{1}=\sqrt{k}$ and $\omega_{2}=\sqrt{k+2k^{\prime}}$.

Continuing with the general analysis, the associated Berry connection can be obtained directly from Eq.~(\ref{Wigner-BerryConn}) with  Eq.~(\ref{Nsco:Atilde}) and it turns out to be
\begin{eqnarray}\label{Nosc-BerryC}
	A_i^{(n_1,\dots,n_N)}(x)=0,
\end{eqnarray}
where we have used the change of variables (\ref{Nosc:transf}) and
\begin{subequations}
	\begin{eqnarray}
	&\int_{-\infty}^{\infty}  {\rm d}Q_a \, {\rm d}P_a\, W_{n_a}=1,\\
	&	\int_{-\infty}^{\infty}  {\rm d}Q_a \, {\rm d}P_a\, f_{n_a\pm 1,n_a}=0, \label{int-fn1} \\
	&\int_{-\infty}^{\infty}  {\rm d}Q_a \, {\rm d}P_a\, f_{n_a\pm 2,n_a}=0. \label{int-fn2}
	\end{eqnarray}
\end{subequations}
Clearly, Eq.~(\ref{Nosc-BerryC}) entails the vanishing of the corresponding Berry curvature
\begin{eqnarray}\label{Nosc-BerryF}
F_{ij}^{(n_1,\dots,n_N)}(x)=0.
\end{eqnarray}
This result is nontrivial and indicates that, for those systems described by a Hamiltonian of the form (\ref{Nosc:quaH}), the (Abelian) Berry phase is zero for arbitrary quantum numbers $n_1,\dots,n_N$ and nondegenerate eigenvalues.

Now we move to the calculation of the Abelian quantum metric tensor. The derivative of the Wigner function~(\ref{Nsco:wigner}) with respect to the parameters leads to
\begin{eqnarray}\label{Nsco:dwigner}
\fl	(\partial_i W_{n_1,n_2})_{q,p} &&= \frac{2}{(2\pi \hbar)^N}{\rm Im}(\mathcal{A}_i^{(n_1,n_2)}) \nonumber\\
\fl	&&=   \sum_{a=1}^{N} \frac{(-1)^{n_a+1}}{2\pi \hbar} {\rm e}^{-\lambda_a/2}  \bigg[ L_{n_a}(\lambda_a) \!-\! 2 \frac{\partial }{\partial \lambda_a} L_{n_a}(\lambda_a) \bigg]  (\partial_i \lambda_a)_{q,p} \prod_{\substack{b=1, \\ b\neq a}}^{N} W_{n_b} ,
\end{eqnarray}
where
\begin{eqnarray}
\fl	(\partial_i \lambda_a)_{q,p} = -\frac{2}{\hbar\omega_{a}} \bigg[ (P_a^2\!-\!\omega_{a}^2 Q_a^2)\frac{\partial_i \omega_{a}}{\omega_{a}}  -2 \sum_{b,c=1}^{N} \left( P_a P_c +\omega_{a}^2 Q_a Q_c \right) U_{cb} \partial_i U_{ab}  \bigg].
\end{eqnarray}
After substituting this result into Eq.~(\ref{qmt-wignerW}), and bearing in mind that
\begin{subequations}
	\begin{eqnarray}
		&&\int_{0}^{\infty}  {\rm d}\lambda_a {\rm e}^{-\lambda_a} (L_{n_a}(\lambda_a))^2 = 1, \label{intL1}\\
		&&\int_{0}^{\infty}  {\rm d}\lambda_a {\rm e}^{-\lambda_a} \lambda_a^2  \left(L_{n_a}(\lambda_a)-2\frac{\partial L_{n_a}(\lambda_a)}{\partial \lambda_a} \right)^2  =  2 (n_a^2+n_a+1) , \label{intL2}\\
		&&\int_{0}^{\infty}  {\rm d}\lambda_a \, {\rm e}^{-\lambda_a} \, \lambda_a (L_{n_a}\,(\lambda_a))^2 = 2n_a+1, \label{intL3}\\
		&&\int_{0}^{\infty}  {\rm d}\lambda_a\, {\rm e}^{-\lambda_a} \lambda_a  \left(L_{n_a}(\lambda_a)-2\frac{\partial L_{n_a}(\lambda_a)}{\partial \lambda_a} \right)^2  = 2n_a+1, \label{intL4}\\
		&&\int_{0}^{\infty}  {\rm d}\lambda_a\, {\rm e}^{-\lambda_a} \lambda_a \, L_{n_a}(\lambda_a) \left(L_{n_a}(\lambda_a)-2\frac{\partial L_{n_a}(\lambda_a)}{\partial \lambda_a} \right) = 1, \label{intL5}
	\end{eqnarray}
\end{subequations}
we arrive at the components of the Abelian quantum metric tensor associated to the family of $N$ coupled harmonic oscillators described by the Hamiltonian~(\ref{Nosc:quaH}):
\begin{eqnarray}\label{Nosc-qmetric}
\fl	g_{ij}^{(n_1,\dots,n_N)}(x) &=&\frac{1}{8} \sum_{a=1}^{N} (n_a^2+n_a+1) \frac{\partial_i \omega_a \partial_j \omega_a}{\omega_a^2}  - \frac{1}{4}  \sum_{a,b=1}^{N} \partial_i U_{ab} \, \partial_j U_{ab} \nonumber\\
\fl	&&+\frac{1}{2}  \sum_{a,b,c,d=1}^{N} \left(n_a+\frac12\right) \! \left(n_b+\frac12\right) \! \left( \frac{\omega_a}{\omega_b} \! + \! \frac{\omega_b}{\omega_a} \right) U_{ac} U_{ad} \partial_i U_{bc} \partial_j U_{bd}.
\end{eqnarray}

This result is new and highly nontrivial. It reveals that the  quantum metric tensor, for the ground state or any excited state, manifests a singular behavior at those points of the parameter space where at least one frequency vanishes. For the ground state of the system, these singularities might correspond to quantum phase transitions~\cite{SHI-JIAN2010}, whereas for excited states,  they might be an indicator of excited-state quantum phase transitions, which can be regarded as extensions of quantum phase transitions~\cite{CAPRIO20081106}. Furthermore, given its simplicity, this metric appears to be suitable for analyzing its large $N$ behavior,  in order to gain a better understanding of the the quantum metric tensor in quantum field theory. On the other hand, notice that for systems where the matrix $U$ does not depend on the parameters, only the terms involving the derivatives of the frequencies in the quantum metric tensor~(\ref{Nosc-qmetric}) survive. Indeed, it is direct to see that for the particular case of the two symmetric coupled harmonic oscillators~(\ref{sco:Hamil}), where $N=2$ and  $\partial_i U_{ab}=0$,  the metric tensor~(\ref{Nosc-qmetric})  reduces to
\begin{eqnarray}\label{sco:metric}
g^{(n_1,n_2)}_{ij}(x)=(n_1^{2}+n_1+1)\frac{\partial_i \omega_1 \partial_j \omega_1}{8\omega_1^2} + (n_2^{2}+n_2+1) \frac{\partial_i \omega_2 \partial_j \omega_2}{8\omega_2^2},
\end{eqnarray}
with $\omega_{1}=\sqrt{k}$ and $\omega_{2}=\sqrt{k+2k^{\prime}}$. This expression shows that the metric is composed of two contributions representing the two uncoupled oscillators, and that its parameter structure is not modified by the quantum numbers. For the particular case $n_1=0$ and $n_2=0$, Eq.~(\ref{sco:metric}) reduces to the quantum metric tensor obtained in Ref.~\cite{AGGV2019} through the path integral approach.

\subsection{Classical analog of the quantum metric tensor}

We would like now to compare and contrast the quantum metric tensor (\ref{Nosc-qmetric}) with its classical counterpart. It is worth recalling that for a classical integrable system described by a Hamiltonian $H(q,p;x)$ with adiabatic parameters $x$, the classical analog of the quantum metric tensor is a metric that provides a measure of the distance between two points in phase space with infinitesimally different parameters. This metric is defined as~\cite{GGVmetric2019}
\begin{eqnarray}\label{classmetric}
	g_{ij}(I,x)=\left< G_i G_j\right>_{\rm Class} - \left<G_i\right>_{\rm Class} \left<G_j\right>_{\rm Class},  
\end{eqnarray}
where $G_i \delta x^i$ is the generator of the infinitesimal canonical transformation $(q(x),p(x)) \rightarrow (q(x+\delta x),p(x+\delta x))$, which in terms of the angle-action variables $(\varphi,I)$ reads
\begin{eqnarray}\label{functionG}
	G_i(\varphi,I;x) := \sum_{a=1}^{N} p_a ( \partial_i q_a)_{\varphi,I} - ( \partial_i S )_{\varphi,I},
\end{eqnarray} 
with $p_a=p_a(\varphi,I;x)$, $q^a=q^a(\varphi,I;x)$, and $S$ the generating function of the canonical transformation $(q,p) \rightarrow (\varphi,I)$. Furthermore, $\left< f(\varphi,I;x)  \right>_{\rm Class}=\frac{1}{(2 \pi)^{n}}\oint d\varphi f(\varphi,I;x)$, with $\oint d\varphi =\prod_{a=1}^{n} \int_{0}^{2 \pi}d\varphi_a$, is the average of $f(\varphi,I;x)$ over the (fast) angle variables.

The classical analog of the quantum Hamiltonian~(\ref{Nosc:quaH}) will be taken to be
\begin{eqnarray}\label{Nosc:classH}
	H=\frac{1}{2} \sum_{a=1}^{N} p_a^2 + \frac{1}{2} \sum_{a,b=1}^{N} K_{ab} q_a q_b, 
\end{eqnarray}
which can be written as $H=(1/2)\sum_{a=1}^{N}(P_{a}^{2}  + \omega_{a}^{2} Q_{a}^{2} )$, after using the linear canonical transformation analogous to that of Eq.~(\ref{Nosc:transf}), i.e., $Q_a=\sum_{b=1}^{N} U_{ab} q_b $ and $P_a=\sum_{b=1}^{N} U_{ab} p_b$, with the generator \begin{eqnarray}
	F=\sum_{a,b=1}^{N} U_{ab} P_a q_b. 
\end{eqnarray}
Notice that the matrix $U$ is the same as in Eq.~(\ref{Nosc:transf}). In turn, the transformation from the variables $(Q,P)$ to the action-angle variables $(\varphi,I)$ is 
\begin{eqnarray}
	Q_a=\left(\frac{2I_a}{\omega_a}\right)^{1/2}\sin\varphi_a, \qquad P_a=\left( 2 \omega_a I_a \right)^{1/2} \cos\varphi_a,
\end{eqnarray}
and its associated generating function is given by
\begin{eqnarray}
	S'=\sum_{a=1}^{N} I_a \left( \varphi_a + \sin\varphi_a \cos\varphi_a \right).
\end{eqnarray}

It is not hard to show that, after the canonical transformations $(q,p) \rightarrow (Q,P) \rightarrow  (\varphi,I)$, the generating function~(\ref{functionG}) takes the form
\begin{eqnarray}\label{class-G}
	G_i= \sum_{a=1}^{N} P_a ( \partial_i Q_a)_{\varphi,I} - ( \partial_i S' )_{\varphi,I} - (\partial_i F)_{q,P}.
\end{eqnarray} 
Then, plugging Eq.~(\ref{class-G}) into Eq.~(\ref{classmetric}) and performing the integrals over the variables $\varphi_a$, the classical analog of the quantum metric tensor is then found to be
\begin{eqnarray}\label{Nosc-classmetric}
 \fl	g_{ij}(I;x) =\frac{1}{8} \sum_{a=1}^{N} I_a^2 \, \frac{\partial_i \omega_a \partial_j \omega_a}{\omega_a^2}  +\frac{1}{2} \sum_{a,b,c,d=1}^{N} I_a I_b \left( \frac{\omega_a}{\omega_b} + \frac{\omega_b}{\omega_a} \right) U_{ac}U_{ad} \partial_i U_{bc} \partial_j U_{bd}.
\end{eqnarray}

In order to compare Eqs.~(\ref{Nosc-qmetric}) and (\ref{Nosc-classmetric}), it is reasonable to use the Bohr-Sommerfeld quantization rule $I_a=(n_a+1/2)\hbar$ and the identifications $I_a^2=(n_a^2+n_a+1)\hbar^2$. Under this consideration, the first thing we notice is that the classical metric~(\ref{Nosc-classmetric}) and the quantum metric~(\ref{Nosc-qmetric}) are identical except for the term $\frac{1}{4}  \sum_{a,b=1}^{N} \partial_i U_{ab} \, \partial_j U_{ab}$ of Eq.~(\ref{Nosc-qmetric}). Actually, we can see that the relation between the two metrics is
\begin{eqnarray}\label{semiclassiacl}
	g_{ij}^{(n_1,\dots,n_N)}(x)=\frac{1}{\hbar^{2}} \bigg( g_{ij}(I;x)- \frac{\hbar^2}{4}  \sum_{a,b=1}^{N}  \partial_i U_{ab}  \partial_j U_{ab}\bigg).
\end{eqnarray}
Clearly, for those systems in which $\partial_i \partial_j U_{ab}=0$, the second term on the right hand side of Eq.~(\ref{semiclassiacl}) vanishes, and then both metrics produce the same parameter structure, satisfying the semiclassical relation established in Ref.~\cite{AGGV2019}. The origin of the extra term can be attributed to the fact that the quantum metric tensor~(\ref{Nosc-qmetric}) has corrections of order $\hbar^2$ to the classical metric~(\ref{Nosc-classmetric}).

\subsection{Example: Linearly coupled harmonic oscillators}

Using Eq.~(\ref{Nosc-qmetric}) it is quite easy to obtain the quantum metric tensor associated to a quantum system described by a Hamiltonian of the form~(\ref{Nosc:quaH}). Let us now illustrate this with a system for which the corresponding matrix $U$ depends on the parameters. We consider the set of two coupled harmonic oscillators described by the Hamiltonian
\begin{eqnarray}\label{lco:Hamiltonian}
	\hat{\bf{H}}=\frac{1}{2}\left(\hat{\bf{p}}_1^2+\hat{\bf{p}}_2^2+A \hat{\bf{q}}_1^2 + B \hat{\bf{q}}_2^2 + C \hat{\bf{q}}_1 \hat{\bf{q}}_2 \right),
\end{eqnarray}
where $x=\{x^{i}\}=(A,B,C)$ ($i,j,\dots\!=1,2,3$) are the adiabatic parameters, which are assumed to satisfy $A\neq B$. Although this model has been discussed extensively in the literature (see Refs.~\cite{Kim_2005,Paz2008,Makarov2018}, for instance), little research has focused on its parameter space~\cite{AGGV2019}. Moreover, its associated quantum metric tensor is known only for the ground state~\cite{AGGV2019}. Here, as a direct application of Eq.~(\ref{Nosc-qmetric}), we shall obtain this geometrical structure for arbitrary quantum numbers.

By using the linear transformation~(\ref{Nosc:transf}) with the parameter-dependent matrix $U$ given by
\begin{eqnarray}\label{lco:U}
	U(x)=
	\begin{pmatrix}
		\cos\alpha & -\sin\alpha  \\
		\sin\alpha & \cos\alpha 
	\end{pmatrix},
\end{eqnarray}
where $\tan\alpha = \frac{\epsilon}{|\epsilon|} \sqrt{\epsilon^2+1} - \epsilon$ with $\epsilon=\frac{B-A}{C}$, the Hamiltonian (\ref{lco:Hamiltonian}) can be written as Eq.~(\ref{Nosc:quaH2}) with $N=2$ and the frequencies
\begin{eqnarray}\label{lco:freq}
	\omega_1=\sqrt{A-\frac{C}{2}\tan\alpha} \qquad  {\rm and} \qquad \omega_2=\sqrt{B+\frac{C}{2}\tan\alpha}.
\end{eqnarray}

Having Eqs.~(\ref{lco:U}) and (\ref{lco:freq}), the computation of the quantum metric tensor~(\ref{Nosc-qmetric}) is a simple exercise in differentiation. We obtain the metric
\begin{eqnarray}\label{lco:qmetric}
\fl	g_{ij}^{(n_1,n_2)}(x)&=& (n_1^{2}+n_1+1)\frac{\partial_i \omega_1 \partial_j \omega_1}{8\omega_1^2} + (n_2^{2}+n_2+1) \frac{\partial_i \omega_2 \partial_j \omega_2}{8\omega_2^2} \nonumber \\
\fl	&&+\left[ \left(n_1 +\frac12\right)  \left(n_2+\frac12\right)  \left(\frac{\omega_1}{\omega_2} + \frac{\omega_2}{\omega_1} \right)- \frac12\right]   \partial_i \alpha \partial_j \alpha ,
\end{eqnarray}
which has the determinant
\begin{eqnarray}
\fl && 	\det[g^{(n_1,n_2)}_{ij}(x)]=\frac{(n_1^{2}+n_1+1)(n_2^{2}+n_2+1)}{4096\omega_{1}^{4} \omega_{2}^{4} (\omega_{1}^{2}-\omega_{2}^{2})^2} \left[ \left(n_1 +\frac12\right) \! \left(n_2+\frac12\right) \! \left(\frac{\omega_1}{\omega_2} + \frac{\omega_2}{\omega_1} \right) \!-\! \frac12\right]\! . \nonumber\\
\fl&&
\end{eqnarray}
It is not hard to see that for the particular case of $n_1=0$ and $n_2=0$, the metric~(\ref{lco:qmetric}) reduces to the one obtained in Ref.~\cite{AGGV2019} by using the path integral formalism. This result then corroborates the validity and  usefulness of Eq.~(\ref{Nosc-qmetric}).

\section{Non-Abelian quantum metric tensor for three coupled oscillators}\label{NonAbelian}

Let us consider a quantum mechanical system composed  of three coupled oscillators and described by the Hamiltonian
\begin{eqnarray}\label{3osc:H}
\fl	\hat{\bf{H}}=\frac{1}{2} \left\{ \sum_{a=1}^{3} \left(\hat{\bf{p}}_a^2 + k \hat{\bf{q}}_a^2\right) + k^{\prime} \big[ (\hat{\bf{q}}_{1}-\hat{\bf{q}}_{2})^{2}  + (\hat{\bf{q}}_{2}-\hat{\bf{q}}_{3})^{2} + (\hat{\bf{q}}_{3}-\hat{\bf{q}}_{1})^{2}\big]\right\}, 
\end{eqnarray}
where $x=\{x^{i}\}=(k,k^{\prime})$ with $i,j,\dots\!=1,2$ are the adiabatic parameters. Note that because this Hamiltonian belongs to the class of quadratic Hamiltonians given by Eq.~(\ref{Nosc:quaH}), we can apply Eq.~(\ref{Nosc-qmetric}) to easily obtain the Abelian quantum metric tensor associated with its nondegenerate ground state. Nevertheless, the aim here is to obtain the Non-Abelian quantum metric tensor for a particular set of degenerate states by means of Eq.~(\ref{Wigner-nA-QMT2}).

Bearing in mind the linear transformation~(\ref{Nosc:transf}) with $N=3$ and the parameter-independent matrix
\begin{eqnarray}\label{3os:U}
	U=
	\begin{pmatrix}
		\frac{1}{\sqrt{3}} & \frac{1}{\sqrt{3}}  & \frac{1}{\sqrt{3}} \\
		-\frac{1}{\sqrt{2}}  & 0 & \frac{1}{\sqrt{2}} \\
		\frac{1}{\sqrt{6}} & -\sqrt{\frac{2}{3}} & \frac{1}{\sqrt{6}} 
	\end{pmatrix},
\end{eqnarray}
the Hamiltonian~(\ref{3osc:H}) can be put in the form
\begin{eqnarray}\label{3osc:quaH2}
	\hat{\bf{H}}=\frac{1}{2} \sum_{a=1}^{3} \left( \hat{\bf{P}}_a^2 + \omega_a^2 \hat{\bf{Q}}_a^2 \right),
\end{eqnarray}
with the frequencies $\omega_{1}=\sqrt{k}$ and $\omega_{2}=\omega_{3}=\sqrt{k+3k^{\prime}}$. Consequently, the normalized wave functions of the system can be written as
\begin{eqnarray}\label{3os:wavef}
	\psi_{n_1,n_2,n_3}(q_1,q_2,q_3;x)=\psi_{n_1}(Q_1;x) \psi_{n_2}(Q_2;x) \psi_{n_3}(Q_3;x),
\end{eqnarray}
where $\psi_{n_a}(Q_a;x)$ is the wave function of the $a$th  uncoupled oscillator with quantum number $n_a=0,1,2,\dots$ and has the same expression as in Eq.~(\ref{sco:wavea}). The energy eigenvalues, which depend on three quantum numbers $n_1$, $n_2$ and $n_3$, are then given by
\begin{eqnarray}
	E_{n_1,n_2,n_3}=\left(n_1+\frac12\right)\hbar \omega_{1} + \left(n_2 +n_3+1\right)\hbar \omega_{2}.
\end{eqnarray}	
Here, we consider the wave functions  $\psi_{0,0,1}$ and $\psi_{0,1,0}$, which have the same energy
\begin{eqnarray}
	E_1:=\frac12 \hbar \omega_{1} + 2\hbar \omega_{2},
\end{eqnarray}	
and then constitute a degenerate set ($g_1=2$). Notice that this degeneracy is a consequence of the fact that the Hamiltonian~(\ref{3osc:quaH2}) is invariant under the interchange of $Q_1$ and $Q_2$.

By introducing the notation $\{\psi_{(1)I}\}:=(\psi_{0,0,1},\psi_{0,1,0})$ with $I,J=1,2$, the associated non-diagonal Wigner functions are obtained by using Eq.~(\ref{nA-f}), which takes the form
\begin{eqnarray}\label{3os-W}
	W_{(1)IJ}=\frac{1}{(2\pi \hbar)^3} \int_{-\infty}^{\infty} {\rm d}^3y \,  {\rm e}^{ -\frac{{\rm i}p\cdot y}{\hbar}} \psi_{(1)I}(q+\tfrac{y}{2};x)\psi_{(1)J}(q-\tfrac{y}{2};x),
\end{eqnarray}
and leads to
\begin{subequations}
	\begin{eqnarray}
		&&W_{(1)11}=\frac{1}{(\pi \hbar)^3} (\lambda_3-1) {\rm e}^{-\frac{\lambda_1+\lambda_2+\lambda_3}{2}},\\
		&&W_{(1)12}=\frac{2}{\pi^3 \hbar^4 \omega_{2}}  (P_2-\rmi \omega_{2} Q_2) (P_3+\rmi \omega_{2} Q_3) {\rm e}^{-\frac{\lambda_1+\lambda_2+\lambda_3}{2}},\\
		&&W_{(1)21}=\frac{2}{\pi^3 \hbar^4 \omega_{2}}  (P_2+\rmi \omega_{2} Q_2) (P_3-\rmi \omega_{2} Q_3) {\rm e}^{-\frac{\lambda_1+\lambda_2+\lambda_3}{2}},\\
		&&W_{(1)22}=\frac{1}{(\pi \hbar)^3} (\lambda_2-1) {\rm e}^{-\frac{\lambda_1+\lambda_2+\lambda_3}{2}},
	\end{eqnarray}
\end{subequations}
where $\lambda_a:=4H_a/\hbar\omega_a$ with $H_a=(1/2)(P_{a}^{2}  + \omega_{a}^{2} Q_{a}^{2} )$.

Then, plugging the expressions for $W_{(1)IJ}$ into  Eq.~(\ref{Wigner-nA-QMT2}), the components of the non-Abelian quantum metric tensor turn out to be
\begin{eqnarray}\label{2os:metric}
	g^{(1)}_{ij11}(x)=&g^{(1)}_{ij22}(x)=\frac{1}{32}\begin{pmatrix}
		\frac{1}{\omega_{1}^{4}}+\frac{4}{\omega_{2}^{4}} & \frac{12}{\omega_{2}^{4}}\\
		\frac{12}{\omega_{2}^{4}} & \frac{36}{\omega_{2}^{4}}
	\end{pmatrix}, \nonumber \\
	g^{(1)}_{ij12}(x)=&g^{(1)}_{ij21}(x)=\begin{pmatrix}
		0 & 0\\
		0 & 0
	\end{pmatrix},
\end{eqnarray}
which are exactly the same that can be obtained directly from Eq.~(\ref{nA-QMT}). Notice that the metric components $g^{(1)}_{ij11}(x)$ and $g^{(1)}_{ij22}(x)$ diverge at the points of the parameter space where the frequencies of the system approach to zero. Besides, note also that $g^{(1)}_{ij11}(x)$ and $g^{(1)}_{ij22}(x)$ can be thought of as invertible matrices, with determinant given by $\det[ g^{(1)}_{ij11}(x)]=\det[ g^{(1)}_{ij22}(x)]=9/256\omega_{1}^{4}\omega_{2}^{4}$. Finally, it is not hard to show that the associated Wilczek-Zee connection and its corresponding curvature are zero. This indicates that non-Abelian quantum metric tensor may contain some valuable information that cannot be extracted from the Wilczek-Zee curvature.

\section{Conclusion}\label{sec:Conclusions}

 In this paper, we have proposed a phase space formulation of the Berry connection, the Wilczek-Zee connection, and the Abelian and non-Abelian quantum geometric tensor, from which the corresponding quantum metrics and curvatures can be obtained in a unified fashion. In the new formulation, these geometrical structures are expressed as integrals over phase space of complex functions, which in principle could be used to obtain additional information about the parameter space. We have shown that the Abelian and non-Abelian quantum metric tensor can be formulated in the phase space formalism by using only parameter derivatives of the diagonal and non-diagonal Wigner functions, respectively. We have also obtained a phase space formulation of the Berry and Wilczek-Zee curvatures, which is different from the one proposed in Ref~\cite{Chruscinski2006}. Indeed, the formulation of Ref~\cite{Chruscinski2006} involves the angle-action variables and therefore, can be applied only to quantum systems whose classical counterpart is integrable. Our formulation, in contrast, uses the new phase space functions $\mathcal{A}_i^{(n)}$~(Abelian case) and $\mathcal{A}_{iIJ}^{(n)}$ (non-Abelian case) and does not have that limitation.
 
 We have illustrated the developed formulation by computing the Berry connection and the quantum geometric tensor for the generalized harmonic oscillator, obtaining the expected results.  As an application of the formulation, we have derived the expression for the Abelian quantum metric tensor associated with a system of $N$ coupled harmonic oscillators. The resulting quantum metric exhibits a singular behavior at the points of the parameter space where there may be a frequency that vanishes. These singularities might be indicators of the presence of quantum phase transitions; however, this is still a matter of on-going research. We have also obtained the classical analog of this Abelian metric and found that both metrics  have the  same parameter structure, except for an extra term that does not involve singularities and that emerges when the transformation that diagonalizes the Hamiltonians depends on the parameters. This result suggests that the classical analog of the quantum metric tensor may help to get a first insight into quantum phase transitions. Finally, we have considered a system of three coupled harmonic oscillators and applied the phase space approach to compute the associated non-Abelian quantum metric tensor, finding that the resulting metric has singularities at the points of the parameter space where the frequencies vanish.
 
 The phase space approach to the parameter space has several advantages over the standard formulation. First, it is suitable to  perform a semiclassical analysis of the geometrical structures involved in the parameter space, because phase space is directly adopted as in many classical concepts. In this line of thought, since the study of chaos is more appropriately carried out in phase space than in configuration space, this approach could be well adapted to seek indicators of quantum chaos in the parameter space framework~\cite{HirschPRL2019,Pandey2020}. Also, this approach seems to be a convenient way to study quadratic Hamiltonians, since it allows a systematic treatment of the system, leading to general results. Furthermore, this approach provides a picture in phase space of what is happening in the parameter space through the associated phase space functions $\mathcal{A}_i^{(n)}$~(Abelian case) and $\mathcal{A}_{iIJ}^{(n)}$ (non-Abelian case), which encode all the information on the underlying  parameter space and serve as the fundamental building blocks for the Abelian and non-Abelian geometrical structures. In addition, and remarkably, in this approach the expression for the Abelian and non-Abelian quantum metric tensor only requires the knowledge of the Wigner functions, which are much closer to classical intuition than the wave functions and can be experimentally measured. In this sense, our procedure also offers an alternative to address the issue of experimentally determining the quantum metric tensor.
 
 The present approach can be used to tackle a variety of conceptual and applied issues. For instance, it can be applied to compute the quantum metric tensor associated with the inverted oscillator, which has been used to study quantum chaos~\cite{Tibra2020} and whose Wigner function is known~\cite{Wolf2010,HEIM20131822}.  Also, as a first step towards understanding the quantum metric tensor in quantum field theory, it would be worth analyzing the large $N$ behavior of the quantum metric tensor obtained in Sec.~\ref{Noscillators} for a lattice of coupled oscillators~\cite{Jefferson2017}. Another important scenario where this approach could be helpful is in analyzing the parameter space associated with coherent states, which have a very convenient representation in the Wigner formalism and have played a central role in quantum optics~\cite{Schleich2001}. Apart from possible applications for pure states, the developed approach may be extended to achieve a phase space formulation of the (non-Abelian) Uhlmann's connection for mixed quantum states~\cite{Uhlmann1986,Uhlmann1989}. Such a formulation would take advantage of a linearity of mixed states in phase space~\cite{Case2008}. Among other applications, the present approach to the parameter space can also be applied to other phase space representations of quantum mechanics, which have their worthiness depending on the kind of application. In particular, it can be implemented in the Husimi representation wherein the state of the system is described by the Husimi function, which is closer to the classical concepts than the Wigner function~\cite{Harriman1993,Fabricio2008} and, for Bloch coherent states, has an appealing interpretation in connection with the Fubini–Study distance~\cite{BookBengtsson}.

\ack
Daniel Guti\'errez-Ruiz is supported with a CONACyT Ph.D. scholarship (No. 332577). Diego Gonzalez was partially supported by a DGAPA-UNAM postdoctoral fellowship and by Consejo Nacional de Ciencia y Tecnología (CONACyT), México, Grant No. A1-S-7701. This work was partially supported by DGAPA-PAPIIT Grant No. IN103919.

\appendix

\section{Alternative derivation of Eq.~(\ref{qmt-wigner}).}\label{app-canonical}

First, let us express Eq.~(\ref{QMT}) in a more suitable form. Taking the second partial derivatives of the density operator $\hat{\rho}_n(x)=\ket{n}\bra{n}$ with respect to the parameters, we have 
\begin{eqnarray}
	\partial_i  \partial_j\hat{\rho}_n=\ket{\partial_i  \partial_jn}\bra{n}+\ket{n}\bra{\partial_i  \partial_jn}+\ket{\partial_i  n}\bra{\partial_jn}+\ket{\partial_j  n}\bra{\partial_in},
\end{eqnarray}
In turn, the normalization condition $\braket{n}{n}=1$ implies $\braket{\partial_i  \partial_jn}{n}+\braket{n}{\partial_i  \partial_jn}+\braket{\partial_i  n}{\partial_jn}+\braket{\partial_jn}{\partial_in}=0$, and hence 
\begin{eqnarray}
\fl 	\bra{n}\partial_i  \partial_j\hat{\rho}_n\ket{n}=-\braket{\partial_i n}{\partial_j n} -\braket{\partial_j n}{\partial_i n}+\braket{\partial_i n}{n}\braket{ n}{\partial_j n}x+\braket{\partial_j n}{n}\braket{ n}{\partial_i n}.
\end{eqnarray}
Plugging this result back into Eq.~(\ref{QMT}), the quantum metric  is expressed by the expectation value of $\partial_i  \partial_j\hat{\rho}_n$, namely
\begin{eqnarray}\label{QMT2}
	g^{(n)}_{ij}(x) = -\frac12 \langle \partial_i  \partial_j \hat{\rho}_n \rangle_n.
\end{eqnarray}

This simple fact, together with Eqs.~(\ref{WignerF}) and~(\ref{WignerAvg}), now provides the connection between the quantum metric tensor and the Wigner function. Actually, from Eqs.~(\ref{WignerAvg}) and (\ref{QMT2}), we have
\begin{eqnarray}\label{QMT3}
	g^{(n)}_{ij}(x) = -\frac12 \int_{-\infty}^{\infty}  {\rm d}^Nq \, {\rm d}^Np\, W_n (\partial_i  \partial_j \tilde{\rho}_n)_{q,p}.
\end{eqnarray}
Then, with the help of Eq.~(\ref{WignerF}), Eq.~(\ref{QMT3}) becomes
\begin{eqnarray}\label{qmt-wigner2}
	g^{(n)}_{ij}(x)=-\frac{(2\pi \hbar)^N}{2} \int_{-\infty}^{\infty}  {\rm d}^Nq \, {\rm d}^Np\, W_n (\partial_i \partial_j W_n)_{q,p},
\end{eqnarray}
which is an equivalent and alternative form of the quantum metric tensor in the Wigner function formalism. Finally, integrating Eq.~(\ref{qmt-wigner2}) by parts and using Eq.~(\ref{dWigner2}), we obtain Eq. (\ref{qmt-wigner}). 

\section*{References}
\bibliography{references}

\end{document}